\documentclass[%
 amsmath,amssymb,
 aip,
 apl,
floatfix,
]{revtex4-1} 

\usepackage{graphicx}
\usepackage{dcolumn}
\usepackage{bm}

\begin{document}

\title{Simultaneous quantitative imaging of surface and magnetic forces}

\author{Daniel Forchheimer}
\author{Daniel Platz}
\affiliation{Royal Institute of Technology, Stockholm}

\author{Erik A. Thol\'{e}n}
\affiliation{Intermodulation Products AB, Solna}

\author{David B. Haviland}
\affiliation{Royal Institute of Technology, Stockholm}

\begin{abstract}
We demonstrate quantitative force imaging of long-range magnetic forces simultaneously with near-surface van-der-Waals and contact-mechanics forces using intermodulation atomic force microscopy. Magnetic forces at the 200~pN level are separated from near-surface forces at the 30~nN level. Imaging of these forces is performed in both the contact and non-contact regimes of near-surface interactions.
\end{abstract}

\maketitle

Separating and identifying the various forces acting between a sharp tip and a sample surface has been a long standing challenge for the interpretation of contrast in atomic force microscopy (AFM). One typical example is the separation of near-surface forces, such as the attractive van-der-Waals and repulsive contact forces which dominate a few nanometers from the surface, from magnetic \cite{Hosaka1992,Kim2009} or electrostatic \cite{Hao1991} forces which dominate at larger distances from the surface. To measure these long-range forces a two-pass method is typically used \cite{Giles1993,Porthun1998}, which has the disadvantage of doubled scan time, loss of resolution and limited ability to measure very close to the surface during the second lifted pass, when the feedback is turned off. Recently multifrequency AFM modes have emerged which provide more measurement channels while scanning \cite{Garcia2012}, allowing single-pass imaging of magnetic forces \cite{Li2009,Dietz2011}. Some of these methods have demonstrated the ability to rapidly capture the entire force-distance curve \cite{Durig2000,Stark2002,Legleiter2006,Sahin2007,Forchheimer2012, Platz2013a, Platz2013b} allowing for simultaneous imaging of topography and material properties. In this letter we perform Intermodulation AFM (ImAFM) \cite{Platz2008} with a magnetically coated tip and analyze the data so as to separate the long-range magnetic force from near-surface forces, thus obtaining simultaneous imaging of topography, mechanical and magnetic properties in one scan, with a standard cantilever, at a typical scan speed for dynamic AFM.

\begin{figure}
\includegraphics[]{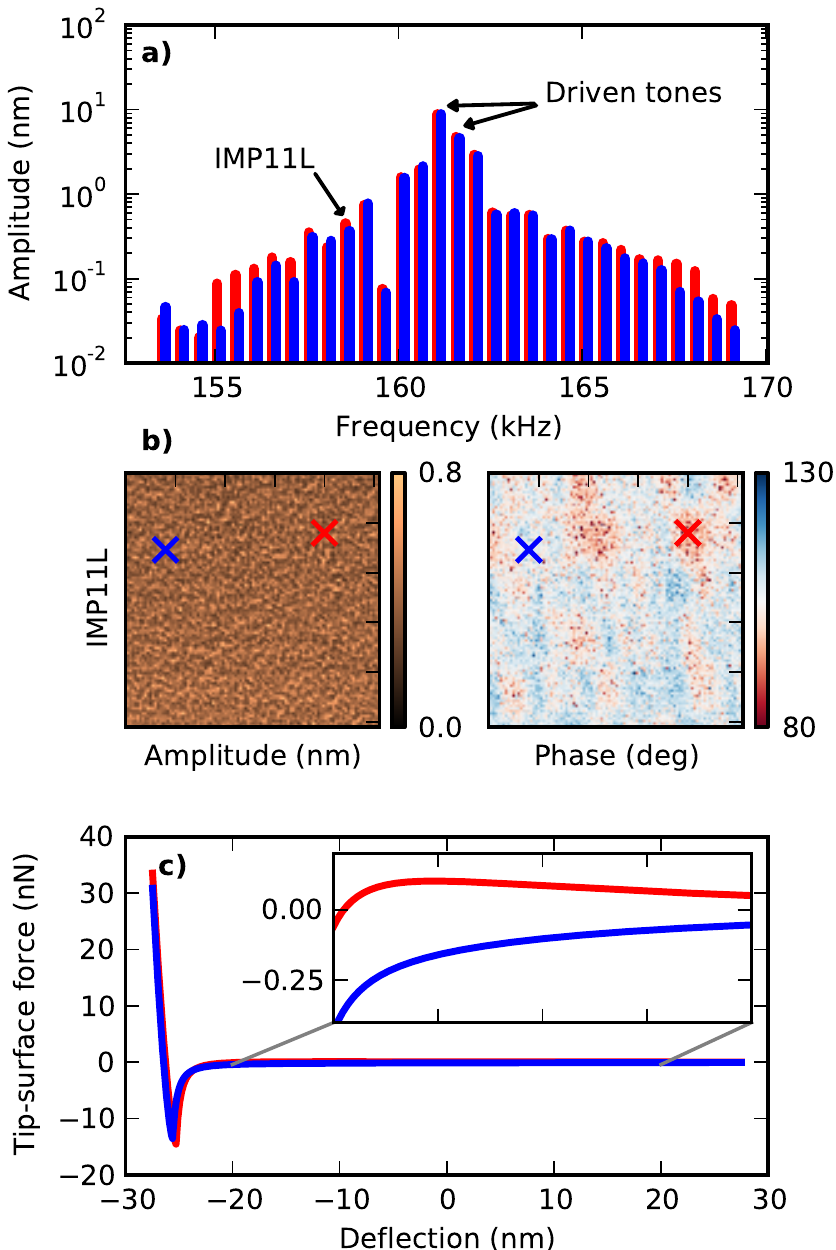}
\caption{(Color online) (a) The amplitudes of two intermodulation AFM spectra (red behind, blue in front) measured at two different positions marked with crosses in (b).  Phases were also measured at each frequency in the spectrum. (b) The amplitude and phase images of the 11th order intermodulation product, for a $1 \mathrm{\mu m}$ scan of the hard disk sample. (c) The reconstructed force-distance curves at the two positions.  The inset shows an expanded vertical scale in the deflection range $-20$ to $20$~nm, where the long-range magnetic force dominates. }
\label{fig:force_distance}
\end{figure}

In narrow-band ImAFM  the cantilever is excited simultaneously with two drive tones closely separated in frequency and centered around a resonance frequency of the cantilever. Upon interaction with the surface the nonlinear tip-surface force creates a spectrum of many intermodulation products of the two drive tones (figure~\ref{fig:force_distance}a), which can be measured by a phase-sensitive lockin technique with good signal-to-noise ratio.  We have previously shown that this spectrum contains the information necessary to accurately reconstruct the tip-surface force curve  and we have demonstrated several methods of force reconstruction from the raw spectral data \cite{Forchheimer2012,Platz2013a,Platz2013b}. One of these methods involves assuming a parametrized force model for the interaction where the force parameters are determined from the measured spectrum with the help of a numerical solver  \cite{Forchheimer2012}.  Here we performed this model-based force reconstruction method  using a force model which takes into account both near-surface and long-range forces. We image a hard disk sample with perpendicular magnetization and observe that the  parameters of the long-range force capture the expected magnetic image.

It is difficult to derive a simple and exact analytic expression describing the force between the magnetic AFM tip and the hard disk surface \cite{Sarid1994,Hartmann1999}.  A good approximate model should have few parameters and the correct asymptotic behavior: The force should go to zero as the distance between the tip and the sample is increased, and the characteristic length for this decay should be the order of the separation of magnetic poles on the surface. We choose a power-law decay,
\begin{equation}
F_{\mathrm{mag}}(z) = F_0 \left( \frac{\lambda_0}{\lambda_0+z-z_{s}} \right) ^p  \hspace*{1cm}  z >z_s
\label{eqn:fmag}
\end{equation}
where $F_{0}$ is the magnitude of the magnetic force at the position of the surface $z=z_{s}$ and $\lambda_0$ the characteristic decay length. We note that the power $p=2$ corresponds to a monopol-monopol while $p=4$ corresponds to dipole-dipole interaction \cite{Jackson1975,Rasa2002}. An estimate of the characteristic decay length is the size of the individual magnetic bits, $170$~nm $\times$ $23$~nm \cite{LLC2008}. Thus we expect the magnetic force to extend significantly beyond the range of the van-der-Waals force which is typically less than $2$~nm.

To account for the surface forces we used the Derjaguin-Muller-Toporov (DMT) model \cite{DERJAGUIN1975,Garcia2002}. The DMT model describes the attractive force of van-der-Waals interactions and the repulsive force due to Hertzian contact mechanics. The qualitative shape of the DMT force-distance curve coincides with that observed when a polynomial \cite{Platz2013b} or model-free force reconstruction \cite{Platz2013a} is performed on the data. The net force is reconstructed as a function of the cantilever deflection $d=z-h$, where $h$ is the probe height and $z$ the instantaneous position of the tip, both being measured in the laboratory frame. In terms of the cantilever deflection the combined force model reads,
\begin{equation}
F_\mathrm{{TS}}(d)=\left\{ 
\begin{array}{lll}
-\frac{HR}{6(d - d_{s})^{2}} & + F_{0} \left( \frac{\lambda_0}{\lambda_0+d-d_{s}} \right) ^p &  d-d_{s} > a_0\\
-\frac{HR}{6a_0^2} & + F_{0} \left( \frac{\lambda_0}{\lambda_0+d-d_{s}} \right) ^p +\frac{4}{3} E^{*}\sqrt{R} \left(-d+d_{s} \right)^{3/2}
&   d-d_s \leq a_0.
\end{array}\right.
\label{eqn:dmt_lin}
\end{equation} 
Where $H$ is the Hamaker constant, $a_0$ is the intermolecular distance in the van-der-Waals model, $E^*$ is the effective Young's modulus, $R$ the tip radius and $d_{s}$ the distance to the surface measured from the probe height $h$.

ImAFM was performed in ambient conditions using a Dimension 3100 (Veeco) AFM with an additional multifrequency lock-in amplifier\cite{Tholen2011} (Intermodulation Products AB) to apply the drive tones and measure the amplitude and phase at 32 response frequencies. A commercially available magnetic force microscopy cantilever MESP-RC (Bruker) was calibrated using the thermal noise method \cite{Sader1999,Higgins2006}. The resonance frequency was $f_0 = 161.3$~kHz, quality factor $Q=269$ and stiffness $k=5.6$~N/m. The cantilever was driven at $f_1 = 161.1$~kHz and $f_2=161.6$~kHz with a total maximum peak-to-peak amplitude of 100~nm.  The measurement bandwidth $\Delta f = f_2-f_1=0.5$~kHz and the image resolution of 256~$\times$~256 pixels determined a scan speed of 1 line per second.  The nominal tip-radius specified by the manufacturer was $35$~nm.

In the 32 amplitude and phase images obtained by ImAFM we could identify two different types of contrasts. The first contrast showed an irregular granular structure with $30-40$~nm grains, matching  the expected tip radius. We attribute this granular contrast to  variations in the surface topography at a scale similar to, or smaller than the tip radius. The second contrast showed larger structures corresponding to the expected magnetic domains. Specifically the $170$~nm magnetic tracks as well as groups of bits were visible, although it was unclear if single bits along the track could be resolved.   Most of the amplitude and phase images contained a mixture of these two contrasts with one sometimes more dominant in the phase, sometimes more dominant in the amplitude.  In general the magnetic contrast was weak compared to the topographic contrast.  Figure~\ref{fig:force_distance}b shows the amplitude and phase images of the lower 11th order intermodulation product at $158.5$~kHz. In this image the granular structure dominates in the amplitude image while both the magnetic and granular contrast can be seen in the phase image.

When reconstructing the force from the intermodulation spectrum with the model-based method, a low number of free parameters aids the solver in finding a distinct minimum of the error function \cite{Forchheimer2012}. To reduce the number of free parameters we fixed the characteristic decay length $\lambda_0$ and power $p$ of the magnetic force. We systematically investigated parameter ranges of $\lambda_0 = 50 - 500$~nm and $p = 1-4$ and found no significant qualitative difference in the contrast observed in the image of the parameter $F_0$. For $\lambda_0 < 10$~nm there was a significant reduction in the contrast of the $F_0$ image. These observations are consistent with the magnetic force decay length being large compared to the oscillation range of the cantilever. Therefore we fixed the parameter values $\lambda_0 = 100$~nm (average dimension of magnetic bit) and $p=4$ (dipole-dipole interaction) for the remaining analysis. The images of the parameter $a_0$, the intermolecular distance in the van-der-Waals force model, showed very little contrast so we fixed this parameter to its mean value over the scan area $a_0 = 0.8$~nm. Furthermore we fix the tip radius $R=35$~nm.

The intermodulation spectrum was analyzed to determine the free parameters of the force model eq.~(\ref{eqn:dmt_lin}) at each pixel of the image \cite{Forchheimer2012}. The force curves at two pixels marked with crosses in figure~\ref{fig:force_distance}b, are plotted in figure~\ref{fig:force_distance}c. The force is dominated by a repulsive interaction of $\sim 30$~nN at peak indentation and an attractive minima of $\sim 15$~nN. This relatively strong attractive force indicates a blunt tip, as one might expect from an AFM tip with a magnetic coating. The difference in magnetic force between the two points can be seen in the region a few nanometers away from the surface (figure~\ref{fig:force_distance}c inset). This difference is quantified in terms of the parameter $F_{0}=0.2$~nN (red cross) and $F_{0}=-0.2$~nN (blue cross). The reconstructed magnetic force is found to be much weaker than the near-surface forces, consistent with the amplitude and phase images being dominated by the granular contrast. Nevertheless, both forces are detected simultaneously and separated from one another.

With ImAFM the force inversion is preformed on the intermodulation spectral data stored at each pixel of the scan to create an image of a force parameter. In the parameter images related to the near-surface interactions $E^*$, $F_\mathrm{min}$ and $d_s$, no prominent features could be seen in the scan area, as expected with a homogeneous sample. However, The long-range force parameter $F_{0}$ generated an image with features corresponding to the magnetic domains of the sample. The parameter images contain noise and  spatial fluctuations caused by the atomic-scale variations in contact geometry. To reduce these effects we applied a spatial Gaussian filter to smooth the parameter values, with a standard deviation $\sigma=10$~nm, the order of the size of the tip (fig.~\ref{fig:parameter_map}, see supplemental material for unsmoothed images).  The $F_0$ image (fig.~\ref{fig:parameter_map}d) clearly shows the magnetic structure  while the near-surface parameters (fig.~\ref{fig:parameter_map}b and c) show only a granular contrast corresponding the the length scale of the smoothing.  We observe both positive and negative values of $F_0$, each having roughly the same magnitude, consistent with the bits being magnetized perpendicular to the surface.


The granulaity seen together with the magnetic structure in the $F_0$ image (fig.~ \ref{fig:parameter_map}d) is the result of cross-talk from the near-surface forces, which are much stronger than the magnetic forces. To mitigate this effect we performed an ImAFM scan where the free-oscillation amplitude was reduced to $40$~nm peak-to-peak.  Lower stored energy in the reduced-amplitude oscillation resulted in a tip-surface interaction that did not reach the repulsive regime \cite{Santos2012}, thus removing the largest force contribution.  The attractive forces were large enough to allow for stable feedback and scanning.  These non-contact measurements were performed with a different cantilever from the same batch having similar calibration values and the magnetic pattern was much more clear in several of the intermodulation amplitude and phase images. For the force reconstruction we used the model eq.~\ref{eqn:dmt_lin} for the case $d-d_{s} > a_0$ only. The tip radius, magnetic decay length and power were again fixed to $R=35$~nm, $\lambda_0 = 100$~nm and $p=4$ respectively, and the free parameters $H$, $d_{s}$ and $F_0$ were obtain from the fit. The parameter image of $F_{0}$ was much improved (fig.~\ref{fig:parameter_map}f) and the $F_0$ values coincided with that determined at larger oscillation amplitude with a different cantilever. The image of the Hamaker constant $H$ (fig.~\ref{fig:parameter_map}e) however showed significant cross-talk with the magnetic image, which may be the result of magnetic forces very close to the surface that are larger than that predicted by the model. Indeed, our model under-estimates the magnetic force at distances less than the bit-size, where a monopol-monopol interaction is expected.   Further improvement of the magnetic image may be possible by improving the magnetic design of the tip \cite{Belova2012}.

One can also visualize the data as slices of the 3-dimensional force volume\cite{Albers2009}. Figure~\ref{fig:force_volume}a shows the function $F_{TS}(z)$ using the fitted and fixed parameter values, plotted the $x-z$ plane for a vertical slice intersecting the $x-y$ plane along the dashed black line in figure~\ref{fig:parameter_map}d.  The force, measured as a function of deflection $d$, was corrected by the feedback height $h$, so that the vertical scale in fig.~\ref{fig:force_volume}a represents the tip position measured from a fixed reference point in the laboratory frame.  The white region at the bottom of fig.~\ref{fig:force_volume}a is outside the range of tip oscillation, and the blue-red interface corresponds to the location of the surface, or the onset of the repulsive force. Figure~\ref{fig:force_volume}b and c shows slices of the force volume in the $x-y$ plane at different heights, indicated by the two dashed lines in fig~\ref{fig:force_volume}a. Near the surface (fig.~\ref{fig:force_volume}b) the van-der-Waals force dominates and at a larger distance from the surface (fig.~\ref{fig:force_volume}c) the magnetic forces dominate. Two movies showing scans of slices in the both the $x-z$ and $x-y$ planes are available in the supplemental material online at [link].

In conclusion, we demonstrate a novel method to separately image the long-range magnetic forces and near-surface forces in atomic force microscopy.  We simultaneously measured the repulsive contact-mechanics force, attractive van-der-Waals force and magnetic force with a single-pass scan, which, to our knowledge, has not been previously demonstrated.  The method can also be applied to other long-range forces such as electrostatic forces, and it provides a calibrated means of quantitatively determining of the separate forces. A long term goal of magnetic force imaging is to obtain quantitative measurements of 3-dimensional magnetic fields and magnetization at the sample surface. Achieving this goal requires not only quantitative measurement of force, but also well-characterized magnetic tips and suitable magnetic force models.


\begin{figure*}
\includegraphics[]{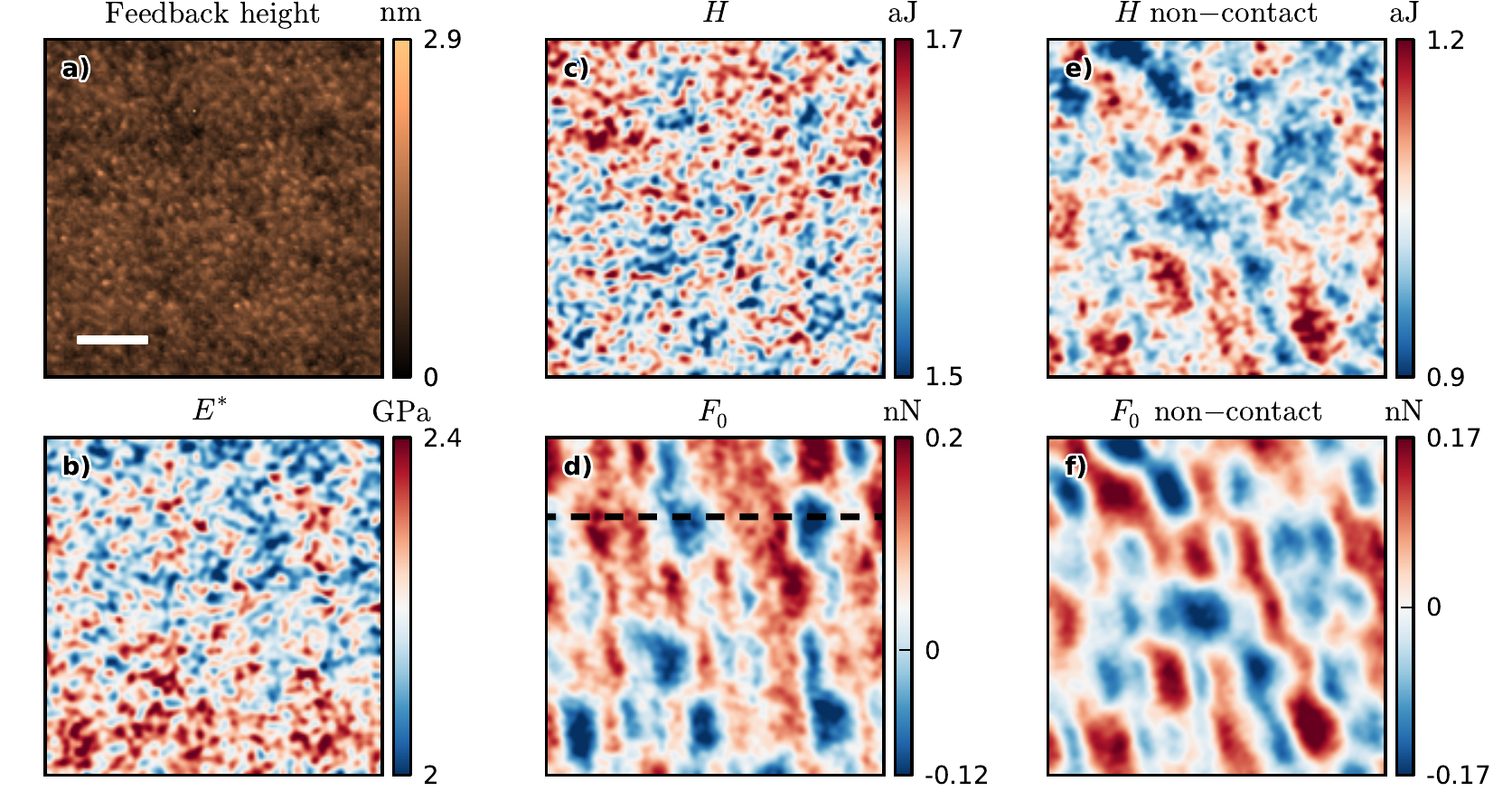}
\caption{(Color online)   Height and parameter images generated from ImAFM scans requiring 4.5 minutes each. The AFM height, or feedback image  (a) shows the surface topography. The white scalebar is 200~nm.   The image of the effective elastic modulus $E^*$(b) and Hamaker constant $H$ (c) show little cross-talk with the magnetic pattern seen in the image of the long-range magnetic force parameter $F_0$(d) . A separate scan in the non-contact regime shows the image of the Hamaker constant $H$ (e) having some cross-talk with  the magnetic force parameter $F_0$ (f).  }
\label{fig:parameter_map}
\end{figure*}

\begin{figure*}
\includegraphics[]{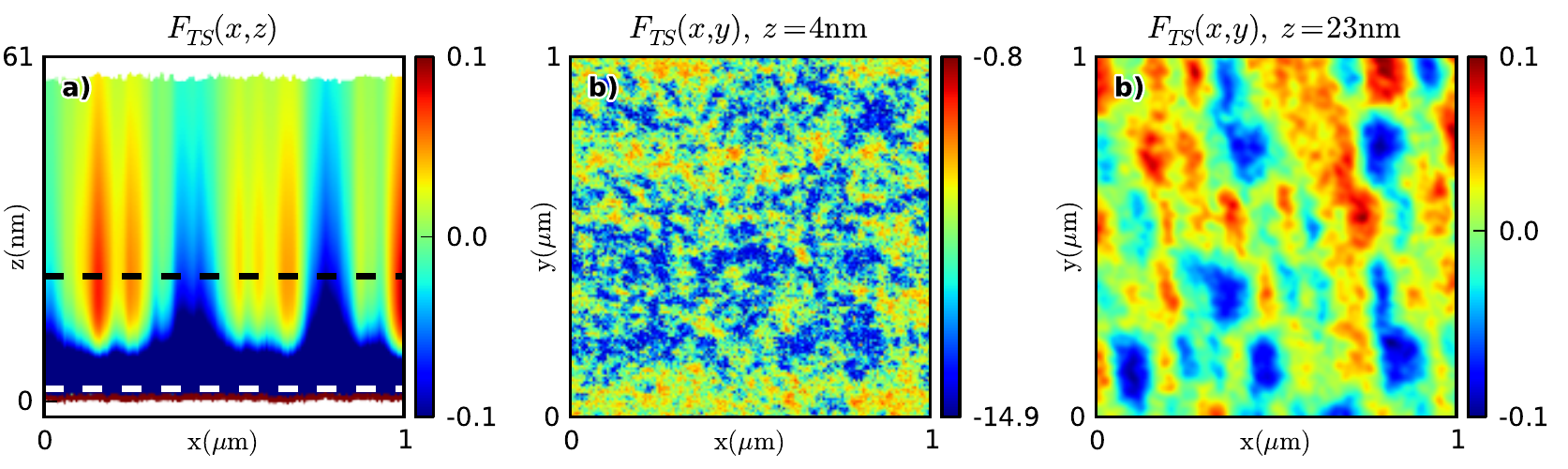}
\caption{(Color online)  Force-volume slices. (a) The force $F_{TS}(x,y,z)$ is displayed in color for a slice in the $x-z$ plane, along the dashed line in \ref{fig:parameter_map}b. The color scale has been limited from -0.1 to 0.1~nN  to enhance the weak long-range magnetic force. (b) The force for a slice in the $x-y$ plane near the surface at $z=$4~nm (white line in (a)) where attractive van-der-Waals force dominate and surface topography is imaged. (c) The force for a slice in the $x-y$ plane away from the surface at $z=23$~nm (black line in (a)0 where magnetic force dominates. Movies showing a scans of both slices are available in the supplemental material online at [link]  }
\label{fig:force_volume}
\end{figure*}

\end{document}